\documentclass[]{spie}  

\usepackage{amsmath,amsfonts,amssymb}
\usepackage{graphicx}
\usepackage[colorlinks=true, allcolors=blue]{hyperref}

\title{Roadmap of GHOST@Gemini’s Precision Radial Velocity Mode}

\author[a]{Venu Kalari}
\author[b]{Andreas Seifahrt}
\author[a]{Ruben Diaz}
\affil[a]{Gemini Observatory/NSF’s NOIRLab, Casilla 603, La Serena, Chile}
\affil[b]{Gemini Observatory/NSF’s NOIRLab, 950 N. Cherry Avenue, Tucson, AZ 85719, USA}

\authorinfo{Further author information: (Send correspondence to V. Kalari)\\V. Kalari: E-mail: venu.kalari@noirlab.edu}

\begin{document} 
\maketitle

\begin{abstract}
GHOST is a newly operational optical fiber-fed high-resolution spectrograph at the Gemini South 8.1m telescope. It currently offers the choice of two resolution modes captured by one (or two) input IFUs with a FOV of 1.2" and a spectral resolving power of 56,000 and 76,000 for the unbinned CCDs. At the high-resolution mode, one can also instigate a simultaneous ThXe calibration lamp, which along with a simultaneous pseudo-slit profile constructed from reformatting the input IFU image will allow for precision radial velocity measurements. Here we talk about the proposed roadmap towards full queue operations, potential upgrades, and the error terms contributing to the final on-sky RV precision, which is estimated to be in the 1-10\,m\,s$^{-1}$ range.
\end{abstract}


\section{GHOST@Gemini}
\label{sec:intro}  

GHOST is a fibre-fed, cross-dispersed, echelle spectrograph located at the Cassegrain focus of the Gemini South telescope. It consists of three main components, namely the bench spectrograph, the Cassegrain unit, and the fiber array, each of which is further subdivided into constituent units.

   \begin{figure} [ht]
   \begin{center}
   \begin{tabular}{c} 
   \includegraphics[height=7cm]{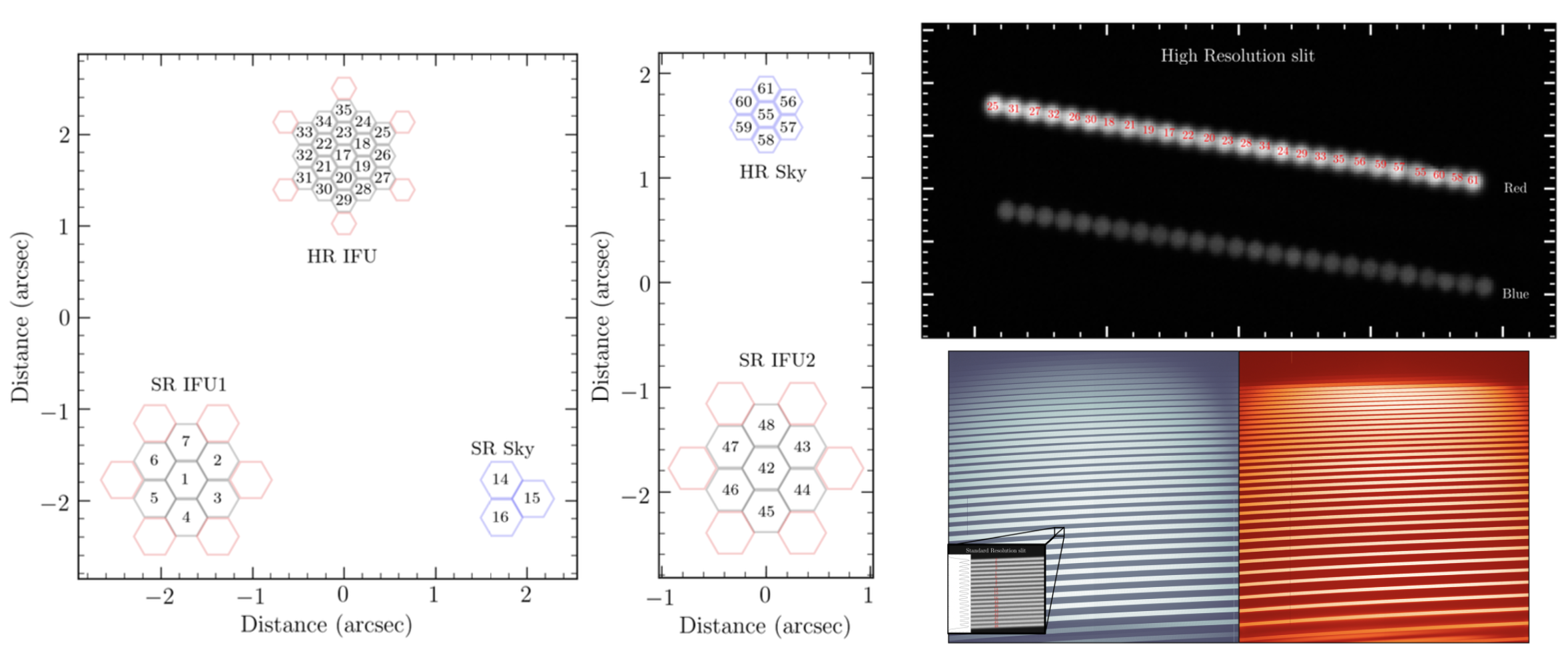}
	\end{tabular}
	\end{center}
   \caption[example] 
   { \label{fig:video-example} 
Overview of the entrance beam in each mode in the left panel, for IFU1 (left) and IFU2 (right), which are then reformatted into a pseudo-slit (top right), before entering the spectrograph (bottom right). In standard resolution mode, the image quality is such that in good conditions, individual microlenses can be resolved (each of 0.4$''$) but not in high resolution mode (0.24$''$). For simultaneous calibration in the precision radial velocity (RV) mode, a ThXe lamp spectra is written below the science orders in the high resolution format. }
   \end{figure}

After light exits the telescope focus, it enters the Cassegrain unit which transmits light from either one, or two individual integral field units (IFUs), each equipped with an internal atmosphere dispersion corrector (ADC). Light is passed then to a bench spectrograph residing in the pier lab, split by a dichroic at 530\,nm into red and blue arms, and captured by individual detectors. An internal ThXe lamp exists to provide simultaneous calibration \cite{kalari, mccon}, and a schematic is given in Fig.\,1.

\section{Current radial velocity precision}

Images at the focal surface are reformatted after exiting the fiber cable system and emerge at a linear pseudoslit (shown in Fig.\,1), increasing the spectral resolution by a factor of 5 (3) in the high (standard) resolution modes. The injection and extraction optics between the on-sky unit to the slit unit by the fiber are the same for the two standard resolution IFUs ($f/$2.8 injection to the fibers), but different for the high-resolution mode (f/4.7 injection to the fibers). For precision radial velocity (PRV) mode, this arrangement ensures the final image entering the slit unit is scrambled in the radial direction, and in the azimuthal direction. As light is injected separately into the spectrograph, the final spectra spectral stability has little dependence on FRD variations. The final pseudo-slit image is captured in the slit unit, which receives 1\% of the light entering the spectrograph. This light further passes through a blue and red filter, covering the 420–600 nm, and the 600–760 nm wavelength range, respectively to capture the bulk of the RV information for early and late-type stars\cite{ireland}. The slit viewing camera obtains simultaneous exposures along with the spectrograph detectors, but can be configured separately allowing for temporal monitoring. 

The spectrograph is in a thermal enclosure and is designed as a box-in-a-box, with active temperature control for all interior and exterior surfaces\cite{pazder, lothrop}. The enclosure is pressure monitored (but not controlled) and is not vacuum enclosed. Shown in Fig.\,2 are the changes in enclosure pressure and grating temperature over the course of this year. Also shown are the changes in the Gaussian full-width half maximum and drift of unresolved high signal lines over the course of two nights in March 2024 compared against the variation in pressure and temperature in the high resolution mode.

\begin{figure} [ht]
\begin{center}
\includegraphics[height=3cm]{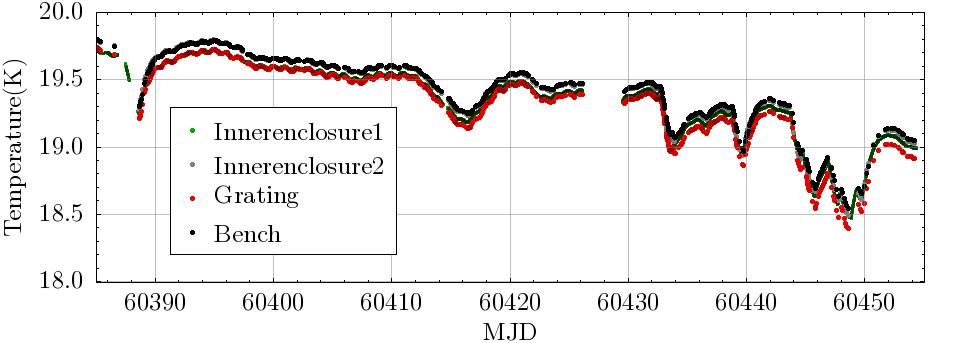}
\includegraphics[height=3cm]{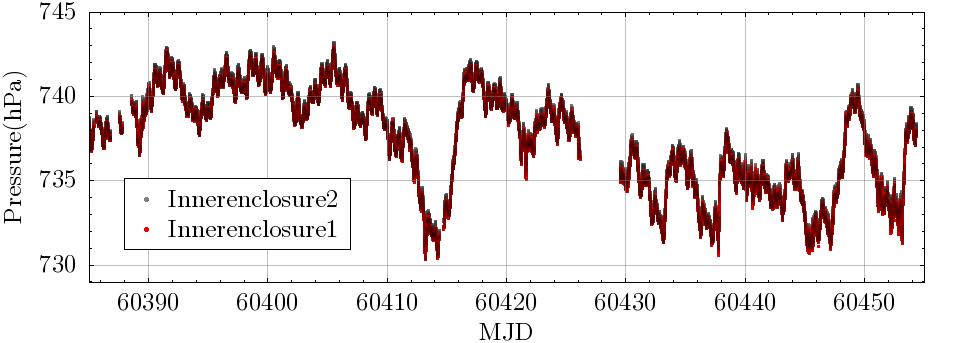}
\end{center}
\caption[example] 
{ \label{fig:video-example} Variation or temperature (left) and pressure (right) at various points in the spectrograph enclosure and the grating over a two month period. }
\end{figure}

\begin{figure} [ht]
\begin{center}
\includegraphics[height=4cm]{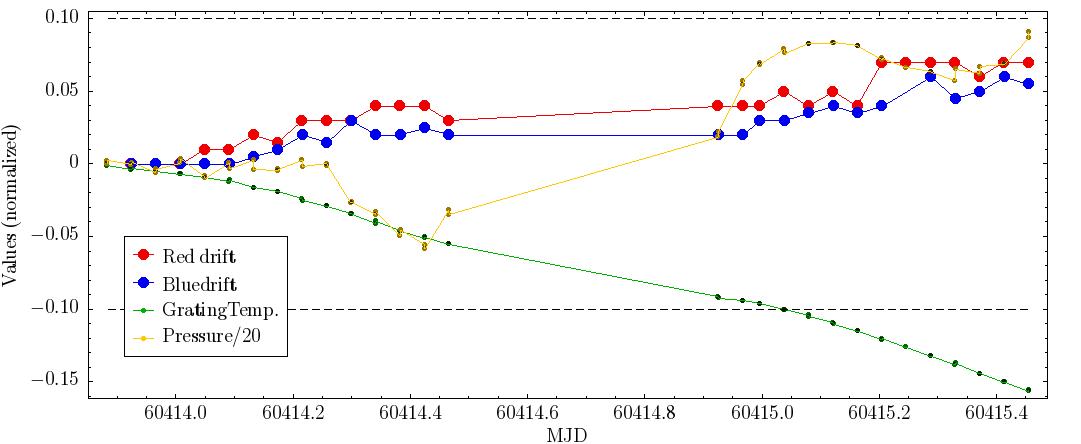}
\end{center}
\caption[example] 
{ \label{fig:video-example} Drift compared to the grating temperature and enclosure pressure over two nights. Dotted lines mark the value of 0.1 pixels from zero. All values are subtracted from the initial value, with pressure divided by a factor of 20 for visual purposes.  }
\end{figure}

The current drift of the spectrograph with changes in pressure and temperature are estimated not to vary within a pixel given the variation in pressure and temperature between the seasonal variations. An example of the RV stability over a shorter period is seen in Fig.\,4 with the RV standard star HD 21693. Observations here achieved an average signal to noise of 200 per resolution element (unbinned spectral data), and observed between airmasses of 1 and 1.4. Results suggest an RV precision over this time period is around 150\,m\,s$^{-1}$ in the standard resolution mode, and around 50\,m\,s$^{-1}$ in the high resolution mode. We expect to achieve higher RV precision and stability, better than 10\,m\,s$^{-1}$ level using the simultaneous calibration accounting for instrumental shifts across the ambient environmental range.

   \begin{figure} [ht]
   \begin{center}
   \begin{tabular}{c} 
   \includegraphics[height=5cm]{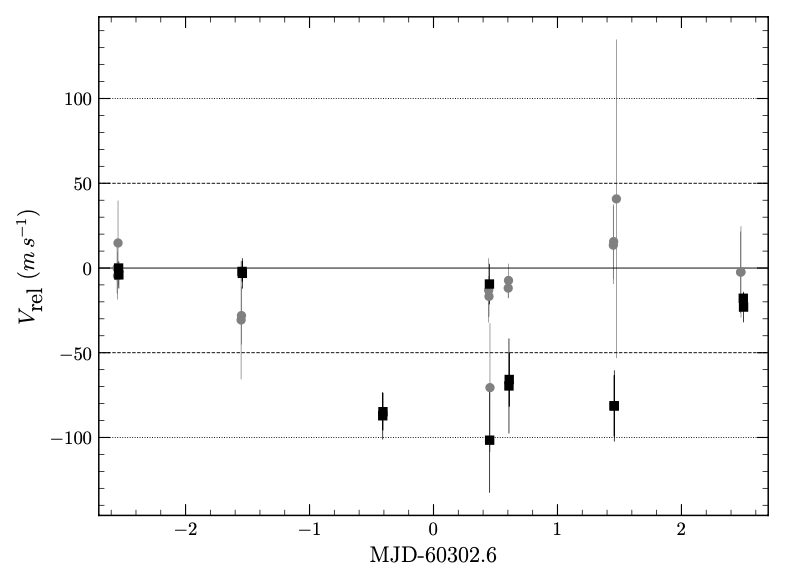}
      \includegraphics[height=6cm]{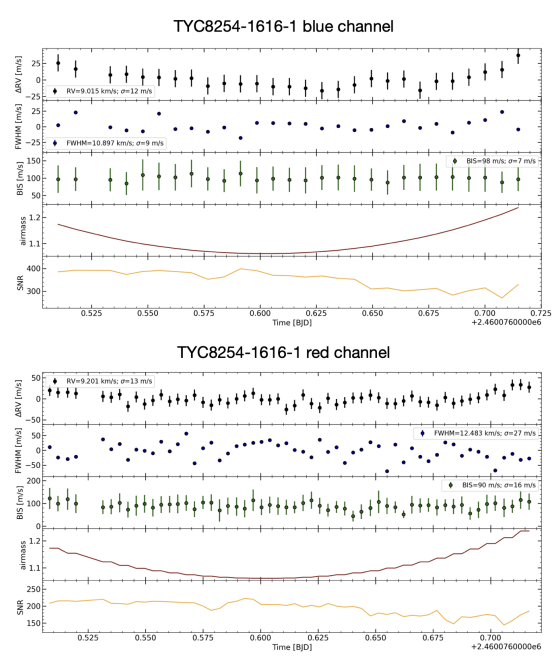}
	\end{tabular}
	\end{center}
   \caption[example] 
   { \label{fig:video-example} 
Relative RV shift measured over a duration of 6 nights on the standard star HD 21693 in both standard (solid black squares), and high resolution mode (light gray circles). The dashed and dotted lines indicate a variation of 50 and 100\,m\,s$^{-1}$, respectively. }
   \end{figure} 

An assessment of the RV performance of GHOST was accomplished during system verification (PI: E. Maritoli). Here, the standard resolution mode of GHOST was used to observe the transit of WASP-108, a hot Jupiter orbiting a F9V star (Fig.\,4). A nearby field star (TYC\,824-1616-1) was observed with the second IFU for comparison. Over the 5 hours of the observation, the RV signal from the Rossiter-McLaughlin effect of the transit was recovered with a rms scatter of roughly 5\,m\,s$^{-1}$, using a classical cross-correlation (CCF) technique against a G2 mask. This result was limited by the absolute drift of the instrument.

\subsection{Known challenges}

\noindent
{\it Instrument profile:} The input slit image produced by the spectrograph is that of the fiber pupils, while the fiber image is enlarged and projected onto the echelle grating. The net result is that the instrumental profile (IP) is largely shaped by the amount of focal ratio degradation (FRD). Any temporal change in FRD, e.g. from varying stresses on the fiber train or small angular misalignments at the IFU, change the IP. IP changes are hard to recover with current hardware.

\noindent
{\it Instrument stability:} Profile variations can arise from minute focus drifts due to temperature, pressure, vibrations, mechanical relaxation, etc. These can not be recovered with standard calibration schemes for fiber-fed spectrographs. These offsets maybe due to IP changes induced by unforeseen environmental factors (observatory wide cooling losses, power losses, etc.). 

\noindent
{\it Input stability:} The slit-viewing camera acts as a exposure meter resolving the incoming flux temporally, spatially, and spectrally. A temporal ‘weighting function’ is constructed using this image to determine the fraction of flux transmitted by each fiber during an integration, and used in the final data reduction and extraction algorithms\cite{ireland14}. Since each fiber has a small (few $\mu$m) offset in the slit and the tolerances of the microlens array forming the pseudo-slit image allows for additional small variations, each fiber will have a net wavelength offset of several 10\,m\,s$^{-1}$. The weighting function derived from the spatially resolved fiber image would compensate for this, but this requires the calibration of the velocity offsets of the fibers in the first place, which is non-trivial given the design of the IFU and the fiber train.

\noindent
{\it Calibration source:} Currently, exposure times that allow recording enough Thorium lines at adequate signal to noise, heavily saturate the Xe lines, particularly in the red arm. Many Xe lines then bleed into the neighboring spectrum of the science fibers. In fact, a large number of lines bleed heavily enough to affect several echelle orders at the same time. A more suitable calibration source, such as Laser Frequency Comb (LFC) or a white light Fabry-Perot etalon is more suitable, and provides a dense comb of flux balanced, unresolved spectral lines.
 
\section{Planned precision radial velocity mode}
\label{sec:sections}

While GHOST has been in queue operations at Gemini since 23B, the PRV mode using simultaneous calibration is yet to be fully commissioned. The stages to do so are outlined here-
{\it Stage 1:} Software control of the internal lamp and filter is complete, allowing to take exploratory science spectra for further testing of the current data reduction pipeline (on-going, data taking to be completed by Aug. 2024).
{\it Stage 2:} An upgraded data reduction pipeline, using the routines for wavelength and drift correction of MAROON-X (Gemini North) is to be adapted for the existing GHOST pipeline (expected Oct. 2024)

Depending on the evaluation of these results, either the mode is released to the community (end of 2024) or the acquisition of further hardware to improve slit calibration, and provide a denser comb of simultaneous calibration spectra begins (mid-2025). A road map of the planned activities scheduled is given in Fig.\,5.

   \begin{figure} [ht]
   \begin{center}
   \begin{tabular}{c} 
   \includegraphics[height=7cm]{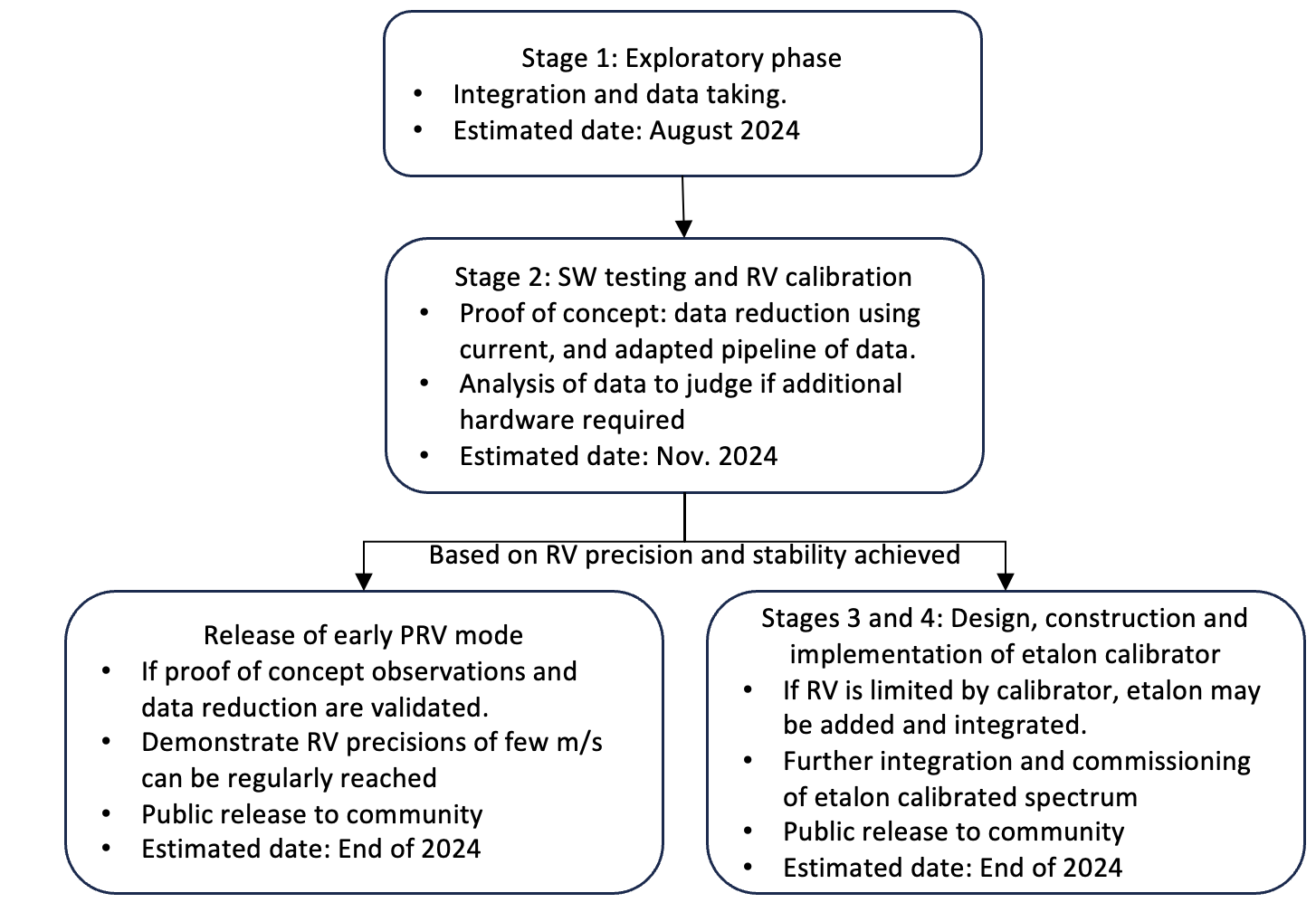}
	\end{tabular}
	\end{center}
   \caption[example] 
   { \label{fig:video-example} 
Roadmap towards operations. }
   \end{figure}

\acknowledgments 
 
We sincerely thank the efforts of the AAO team (lead G. Robertson), ANU team (lead M. Ireland), NRC team (lead A. McConnachie), and the Gemini staff. This work was supported by and based on observations obtained at the international Gemini Observatory, a program of NSF’s NOIRLab, which is managed by the Association of Universities for Research in Astronomy (AURA) under a cooperative agreement with the National Science Foundation on behalf of the Gemini Observatory partnership: the National Science Foundation (United States), National Research Council (Canada), Agencia Nacional de Investigacio n y Desarrollo (Chile), Ministerio de Ciencia, Tecnolog ıa e Innovaci on (Argentina), Minerio da Ciencia, Tecnologia, Inovacoes e Comunicaoes (Brazil), and Korea Astronomy and Space Science Institute (Republic of Korea). 
\bibliography{report} 
\bibliographystyle{spiebib} 

\end{document}